\documentstyle[aps,multicol]{revtex}
\renewcommand{\narrowtext}{\begin{multicols}{2} \global\columnwidth20.5pc}
\renewcommand{\widetext}{\end{multicols} \global\columnwidth42.5pc}
\newcommand{\onetext}{\end{multicols} \global\columnwidth20.5pc}

\input{epsf.tex}
\def\inseps#1#2{\def\epsfsize##1##2{#2##1} \centerline{\epsfbox{#1}}}

\def \attn #1 {{\sl $\bullet$ #1 $\bullet$}}
\begin{document}
\draft

\title{A Composite Fermion Description of Rotating Bose-Einstein
Condensates}
\author{N. R. Cooper and N. K. Wilkin}
\address{School of Physics and Astronomy, University of Birmingham,
Edgbaston, Birmingham, B15 2TT, United Kingdom.}
\date{\today}

\maketitle

\begin{abstract}
We study the properties of rotating Bose-Einstein condensates in
parabolic traps, with coherence length large compared to the system
size.  In this limit, it has been shown that unusual groundstates form
which cannot be understood within a conventional many-vortex
picture. Using comparisons with exact numerical results, we show that
these groundstates can be well-described by a model of non-interacting
``composite fermions''. Our work emphasises the similarities between
the novel states that appear in rotating Bose-Einstein condensates and
incompressible fractional quantum Hall states.
\end{abstract}


\narrowtext


It has proved fruitful in fractional quantum Hall
systems\cite{prangeandgirvin} to account for the many-body
correlations induced by electron-electron interactions by introducing
non-interacting ``composite fermions''\cite{cf}. Recently a similar
approach has been employed to show that the correlated states arising
from interparticle interactions in dilute rotating confined bose
atomic gases can be described in terms of the condensation of a type
of composite {\it boson}\cite{wilkinunpub}.  Here, we demonstrate that
a transformation of the system of rotating bosons to that of
non-interacting composite fermions is also successful in accounting
for these correlated states. Our results establish a close connection
between the groundstates of rotating confined Bose-systems and the
correlated states of fractional quantum Hall
systems\cite{prangeandgirvin}.

While the trapped atom gases have been shown to
Bose-condense\cite{JILA,Dalfovo1999}, the response of these
condensates to rotations has not, as yet, been measured
experimentally.  Theoretically, it is clear that there exist various
different regimes.  Within the Gross-Pitaevskii framework, which
requires macroscopic occupation of the single particle states, the
system forms vortex arrays at both long\cite{Butts1999} and
short\cite{Castinunpub} coherence lengths (compared to the size of the
trap), which are reminiscent of Helium-4. Here, following
Ref.~\onlinecite{wilkinunpub}, we choose to study the system in the
limit of large coherence length without demanding macroscopic
occupation numbers. This allows us to study both the regime considered
in Ref.~\onlinecite{Butts1999}, as well as regimes of higher vortex
density where the quantum mechanical nature of the vortices will be
most prevalent.  Indeed, in Ref.~\onlinecite{wilkinunpub} it was shown
that, in general, the groundstates of the rotating boson system cannot
be described within a conventional many-vortex picture.  Rather, the
system was found to be better described in terms of the condensation
of ``composite bosons'' -- bound states of vortices and atoms --
across the whole range of vortex density.  In the present paper, we
show that a description in terms of non-interacting composite
particles with {\it fermionic} statistics also provides a highly
accurate description of the rotating bose system: specifically, it
enables us to predict many of the features in the energy spectrum and
to form good overlaps with the exact groundstate wavefunctions.  In
addition, this description indicates a close relationship between the
properties of rotating Bose systems and those of fractional quantum
Hall systems.

In a rotating reference frame, the standard Hamiltonian for $N$ weakly
interacting atoms in a trap is\cite{Dalfovo1999}
\begin{equation}
{\cal H} =\frac{1}{2}\sum_{i=1}^N[-\nabla^2_i + r_i^2 + \eta \sum_{j=1,\ne
i}^N\delta(\bbox{r}_i-\bbox{r}_j) - 2\bbox{\omega}\cdot \bbox{L}_i]
\label{eq:ham}
\end{equation}
where we have used the trap energy, $\hbar \sqrt{K/m} = \hbar\omega_0$
as the unit of energy and the extent, $(\hbar^2/MK)^{1/4}$, of the
harmonic oscillator ground state as the unit of length. ($M$ is the
mass of an atom and $K$ the spring constant of the harmonic trap.)
The coupling constant is defined as $\eta = 4\pi {\bar
n}a(\hbar^2/MK)^{-1/2}$ where ${\bar n}$ is the average atomic density
and $a$ the scattering length.  The angular velocity of the trap,
$\omega$, is measured in units of the trap frequency.

Throughout this work, we make use of the limit of weak interactions
($\eta\ll 1$). It was shown in Ref.~\onlinecite{Wilkin1998} that in
this limit the system may be described by a two-dimensional model with
a Hilbert space spanned by the states of the lowest Landau level:
$\psi_m(\bbox{r}) \propto z^m \exp(-zz^*/2)$, where $m$ is the angular
momentum quantum number ($m=0,1,2\ldots$) and $z\equiv x+iy$.  The
kinetic energy is quenched and the groundstate is determined by a
balance between the interaction and potential energies.  Noting that
the $\bbox{z}$-component of the angular momentum, $L$, commutes with
the Hamiltonian, the total energy, scaled by $\eta$, may be written
\begin{equation}
E/\eta = V_N(L) + (1-\omega)/\eta L ,
\label{eq:energy}
\end{equation}
where $V_N(L)$ is the interaction energy at angular momentum $L$.
While this separation holds for all energy eigenstates, we choose
$V_N(L)$ to denote the smallest eigenvalue of the interactions at
angular momentum $L$.  Since the interactions are repulsive, $V_N(L)$
decreases as $L$ increases and the particles spread out in space; a
tendency that is opposed by the term $(1-\omega)/\eta L$ describing
the parabolic confinement.  Thus, as the rotation frequency $\omega$
is varied, the groundstate angular momentum will increase, from $L=0$
at $\omega=0$, to diverge as $\omega\rightarrow 1$ (when the trap
confinement is lost); our goal is to describe the sequence of states
(of different $L$) through which it passes.

\begin{figure} \inseps{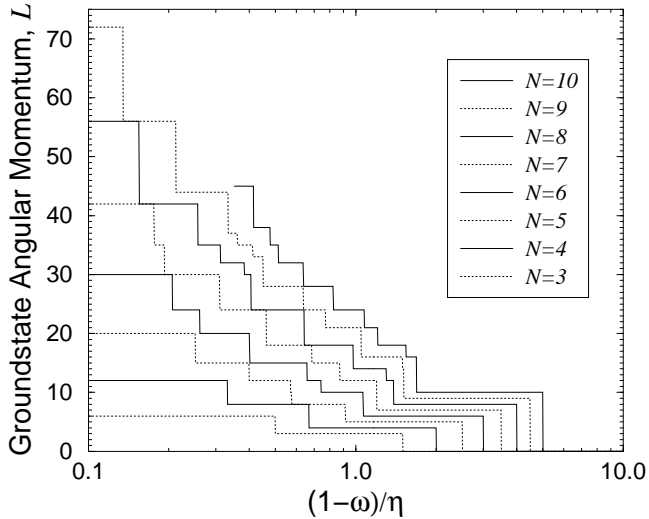}{0.5}
\caption{Groundstate angular momentum as a function of rotation
frequency, $\omega$, for $N=3\rightarrow 10$
particles\protect\cite{lanc}.}
\label{fig:stable_states}
\end{figure}

We have obtained the groundstate interaction energies, $V_N(L)$, for
$N=3$ to $10$ particles, from exact numerical diagonalisations within
the space of bosonic wavefunctions in the lowest Landau
level\cite{lanc}.  While the interaction energy $V_N(L)$ does decrease
with increasing angular momentum, it is not a smooth function of $L$.
Thus, the groundstate angular momentum, obtained by minimising
(\ref{eq:energy}), is not a smoothly increasing function of $\omega$.
As shown in Fig.~\ref{fig:stable_states}, certain values of angular
momentum, corresponding to downward cusps in $V_N(L)$, are
particularly stable, and are selected as the groundstate over a range
of $\omega$.
The existence of certain angular momentum states of enhanced stability
is reminiscent of the ``magic-values'' of angular momentum for
electrons in quantum dots\cite{Lai1984,Maksym,Yang1993,jainkawamura};
by analogy we will also refer to the stable angular momenta of the
bosons as the magic values.  Indeed, the system of bosons we study is
precisely the {\it bosonic} variant of the (fermionic) problem of a
parabolic quantum dot in strong magnetic field, with $\omega_0$
playing the role of the magnetic field and $(1-\omega)/\eta$ the role
of the parabolic confinement, and with $\delta$-function interactions
replacing the more usual Coulomb repulsion.  As we shall explain, the
magic values of angular momenta for the bosonic and fermionic systems
are, in fact, closely related. This is a corollary of our principal
result, to which we now turn, that much of the structure appearing in
Fig.~\ref{fig:stable_states} can be interpreted simply in terms of the
formation of bound states of bosons and vortices behaving as
non-interacting composite particles with {\it fermionic} statistics --
``composite fermions''(CF).


It is known that, for {\it homogeneous systems}, interacting bosons
and interacting fermions within the lowest Landau level have many
features in common.  For example, there exist certain filling
fractions\cite{fillingfraction} of both the boson and fermion systems
at which interactions lead to incompressible groundstates, with
wavefunctions that may be related by a simple statistical
transformation if $1/\nu_F = 1/\nu_B + 1$\cite{Xie1991} ($\nu_B$ and
$\nu_F$ are the filling fractions of the bosons and fermions).  These
similarities arise from the remarkable effectiveness of mean-field
approximations to Chern-Simons theories of such systems\cite{cf}.
Here, we are interested in an {\it inhomogeneous} system, in which the
bosons are subject to a parabolic confinement.  Jain and
co-workers\cite{jainkawamura,kamillajain,kawamurajain} have shown that
the fermionic equivalent of this problem -- interacting electrons in a
quantum dot -- can be well-described in terms of properties of {\it
non-interacting composite-fermions}. Motivated by the successes of
their theory, we apply a similar transformation to describe the
present bosonic problem.

Specifically, we make the following ansatz for the many-boson
wavefunction
\begin{equation}
\Psi^{ansatz}_L(\{z_i\})  =  {\cal P} \left\{ \prod_{i<j} (z_i-z_j)
\Psi^{CF}_{L_{CF}}(\{z_i\}) \right\}
\label{eq:transform}
\end{equation}
where $\Psi^{CF}_{L_{CF}}(\{z_i\})$ is a wavefunction for some
fermionic particles -- the composite fermions. Multiplication of the
antisymmetric CF wavefunction by the Jastrow prefactor generates a
completely symmetric bosonic wavefunction. ${\cal P}$ projects the
wavefunction onto the lowest Landau level, which amounts to the
replacement $z_i^n\bar{z_i}^m \rightarrow \{ \frac{n!}{m!} z_i ^{n-m} \; (n \ge m) \; ; 0 \; (n<m)\}$ for all terms in the polynomial part of the
wavefunction. For a full discussion see
Ref.~\onlinecite{jainprb}. (For ease of presentation, we omit
exponential factors and normalisation constants from all
wavefunctions.)

\begin{figure}
\inseps{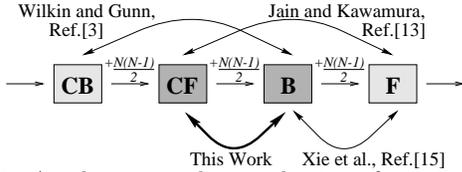}{0.4}
\caption{A schematic relating the transformation we are performing to
those that have been used previously.}\label{fig:relation}
\end{figure}

The transformation (\ref{eq:transform}) causes the boson wavefunction
to describe a half vortex around the position of each other particle
in addition to the motions described by $\Psi^{CF}$. One can therefore
interpret a composite fermion as a bound state of a boson with a half
vortex (cf. Ref.\onlinecite{Read1996}). As a result, the angular
momentum of the bosons, $L$, is increased with respect to that of the
composite-fermions, $L_{CF}$, according to
\begin{equation}
L = L_{CF} + N(N-1)/2 .
\label{eq:angmtm}
\end{equation}
Note that the transformation (\ref{eq:transform}) relates $1/\nu_{CF}
= 1/\nu_B - 1$ and is not the same as that used in
Ref.~\onlinecite{Xie1991}.  There are an unlimited number of fermion
$\leftrightarrow$ boson mappings that one can effect through
transformations of the form (\ref{eq:transform}).  In
Fig.~\ref{fig:relation} we present a schematic of how, by subsequent
attachments of half-vortices -- each causing an addition of $N(N-1)/2$
to the angular momentum -- one can transform from the composite bosons
(CB) introduced in Ref.~\onlinecite{wilkinunpub} to the composite
fermions used here (CF), to the bare boson system in which we are
interested (B), and finally to a fermion system (F).
We introduce the fermion system (F) to point out that the composite
fermions (CF) which we use to describe the boson system (B) are the
same as those used by Jain and Kawamura\cite{jainkawamura} to describe
interacting electrons in quantum dots (F).  The predictions of the
energy spectrum flowing from a model of non-interacting composite
fermions will therefore be identical in the boson and fermion systems
up to the shift $L_{F}=L + N(N-1)/2$.

In the spirit of Ref.~\onlinecite{jainkawamura}, we shall consider the
CFs, described by $\Psi^{CF}_{L_{CF}}(\{z_i\})$, to be
non-interacting, and look at the variation of the minimum {\it
kinetic} energy of the CFs as a function of the total angular
momentum.  We further assume that a composite fermion in the Landau
level state $(n,m)$ (with Landau level index $n=0, 1, 2 \ldots$, and
angular momentum $m=-n,-n+1\ldots$) has an energy $E_n = (n+1/2)
E_{CF}$, where $E_{CF}$ is some effective cyclotron energy.  These
assumptions may be viewed as a mean-field treatment of the appropriate
Chern-Simons theory for this system; ultimately, they are justified by
the predictive successes of the resulting theory.

\begin{figure}
\inseps{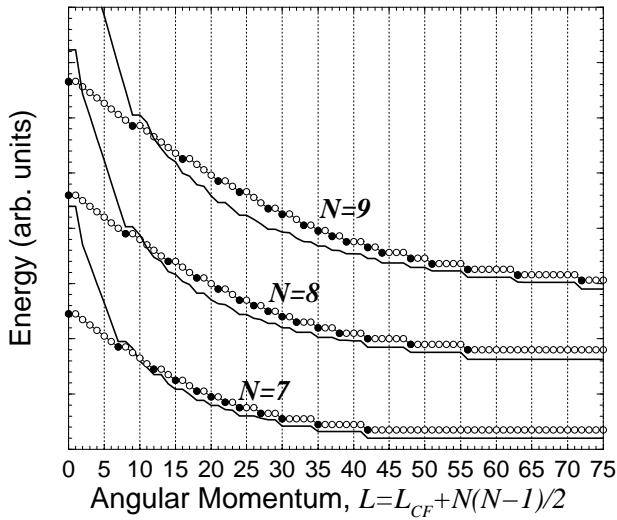}{0.5}
\caption{The circles show the groundstate energies of non-interacting
composite fermions as a function of angular momentum, for $N=7, 8, 9$
(filled circles identify the sets, $L^*_N$, of angular momenta at
which the CF energy has a downward cusp). The exact interaction
energies, $V_N(L)$ are shown as solid lines for comparison. (All
curves are offset for clarity.)}
\label{fig:freecf_energies}
\end{figure}

Figure~\ref{fig:freecf_energies} shows the resulting groundstate
energy of non-interacting CFs as a function of $L = L_{CF} + N(N-1)/2$
for $N=7, 8, 9$, together with the exact interaction energies
$V_N(L)$.
It is apparent that the composite fermion energies fail to capture the
rapid rise in the exact energies at small angular momenta; this can be
interpreted as a failing of the assumption of a constant effective
cyclotron energy $E_{CF}$.  The principal success of this approach is
the identification of the {\it cusps} in the exact energy $V_N(L)$: at
almost all of the angular momenta for which the composite fermion
kinetic energy shows a downward cusp (we label these sets of angular
momenta by $L^*_N$), there is a corresponding cusp in the exact energy
$V_N(L)$. Since a ``magic'' angular momentum of the boson system must
coincide with a downward cusp in $V_N(L)$, the set $L^*_N$ represents
{\it a set of candidate values for the magic angular momenta}.  For
example: for $N=7$, the CF model predicts all nine actual magic
numbers and in addition identifies cusps which do not become
groundstates for a further three $L\in L^*_7$. These missing values
are not necessarily a failing of the composite fermion model. The main
failing, to which we will return later, is that there is a small
number of magic angular momenta that are not identified.

Not only does the composite fermion model successfully identify the
majority of the magic angular momenta, as we now show it also provides
a very accurate description of the associated wavefunctions.  The
composite fermion wavefunctions corresponding to the angular momenta
$L^*_N$ are the ``compact states'' discussed in
Ref.~\onlinecite{jainkawamura}.  For these states, the composite
fermions occupy the lowest available angular momentum states within
each Landau level.  As an illustration, for $N=4$, there is a cusp in
the composite fermion energy at $L=8$ ($L_{CF}=2$), at which the
composite fermions occupy the single particle states $(n,m) =
\{(0,0),(0,1),(0,2),(1,-1)\}$.  The wavefunction $\Psi^{CF}_{L_{CF}}$
is formed as a Slater determinant of these states, and the bosonic
wavefunction $\Psi^{ansatz}_L$ is constructed via
Eq.(\ref{eq:transform}).

In the cases $L=0$ and $L= N(N-1)$, this procedure yields the exact
groundstate wavefunction for all $N$.  At $L=0$, there is only one
many-body state within the lowest Landau level (all bosons occupy the
$m=0$ state); the ansatz (\ref{eq:transform}) has non-zero overlap
with this state, so must (trivially) be the groundstate.  For
$L=N(N-1)$, the lowest energy composite fermion state is formed from
the states $\{(0,0),(0,1),....(0,N)\}$. The Slater determinant of
these states may be written $\Psi^{CF} = \prod_{i<j} (z_i-z_j)$,
which, inserted in (\ref{eq:transform}), generates the bosonic
Laughlin state $\Psi^{ansatz} = \prod_{i<j} (z_i-z_j)^2$.  (This state
is in the lowest Landau level, and projection is unnecessary.)  Since
this wavefunction vanishes for $z_i=z_j$ ($i\neq j$), it is the exact
zero energy eigenstate of the $\delta$-function two-body interaction
potential.

At intermediate values of the angular momentum, our ansatz
(\ref{eq:transform}) is not, in general, exact. We have performed
numerical calculations to determine the overlaps of the ansatz
wavefunctions with the exact groundstate wavefunctions, $|\langle
\Psi^{ansatz}_L | \Psi^{exact}_L\rangle|$.  We list these overlaps in
Table~\ref{table} at each of the angular momenta, $L\in L^*_N$,
selected by the non-interacting composite fermion model.  In general,
the ansatz (\ref{eq:transform}) has an overlap of close to unity with
the exact groundstate: the composite fermion model provides an
excellent description of these states.  Small overlaps can occur when
the composite fermion model does not produce a unique ansatz -- {\em
i.e.} when two, or more, sets of single particle states for the
composite fermions have the same kinetic energy at a given $L$ (e.g.
$N=6, L=12$).  In these cases, the overlaps could be improved by
diagonalising the Hamiltonian within the space of states spanned by
the two ansatz states.

Owing to the impressive agreement between the ansatz wavefunctions
(\ref{eq:transform}) and the exact groundstates at $L^*_N$, an
accurate description of the groundstate angular momentum as a function
of rotation frequency can be obtained using only this set of ansatz
wavefunctions. Minimizing the expectation value of the energy
(\ref{eq:energy}) within this set of ansatz wavefunctions, one obtains
a groundstate angular momentum as a function of $(1-\omega)/\eta$ that
is in excellent agreement with the exact results shown in
Fig.~\ref{fig:stable_states}.  This approach does, however, omit a
small number of magic angular momenta.  In some cases, these are magic
values identified by the composite fermion model, but for which the
expectation value of the energy happens not to be sufficiently low to
become stable ($N=6$, $L=12$; $N=9$, $L=33,37$; $N=10$, $L=38$).  The
most important omissions are the magic angular momenta at $N=8, L=12$,
$N=10, L=16,21$ for which there are {\it no} features in the composite
fermion kinetic energy that would suggest a stable angular momentum
state.  We believe that this emergent structure at larger numbers of
particles represents many-body correlations that are not captured by
the non-interacting composite fermion model used here. (Some of these
states are correctly identified by the composite boson
approach\cite{wilkinunpub}.) They could be related to the
incompressible states, such as $\nu=4/5$, of quantum Hall systems
which cannot be explained in terms of non-interacting composite
fermions alone, but require an additional `particle-hole'
transformation.  This view is strengthened by the observation that
related magic angular momenta also appear in the exact groundstate
energy of electrons in quantum dots interacting by Coulomb forces, up
to the the shift $L_F = L + N(N-1)/2$ ({\em e.g.}
Ref.~\onlinecite{kawamurajain} identifies a stable state of $N=10$
electrons at $L_{F}=61$ -- equivalent $N=10, L=16$ of the present
bosonic model).  The study of this additional structure is beyond the
scope of the present work.


In summary, we have studied the properties of rotating Bose systems in
parabolic traps in the limit of large coherence length.  Through
comparisons with exact results for small systems, we showed that many
of the features of the exact spectrum of the bosons can be understood
in terms of {\it non-interacting composite fermions}.  The
non-interacting composite fermion model leads to (1) the
identification of a set of candidate values for the stable angular
momenta of the bosons, and (2) associated many-body wavefunctions that
have large overlap with the exact groundstate wavefunctions.  The
successes of the mapping to composite fermions indicate that the
groundstates of rotating Bose-Einstein condensates, in the limit of
large coherence length, are closely related to the correlated states
appearing in fractional quantum Hall systems.

\vskip0.1cm We would like to thank J.M.F. Gunn and R.A. Smith for
many helpful discussions. This work was supported by the Royal Society
and EPSRC GR/L28784.

\newpage

\onetext

\begin{table}
\begin{tabular}{|c||l|}
$N$ & Angular Momentum  $L$,  [$|\langle \Psi^{ansatz}_L
|\Psi^{exact}_L\rangle|$]
\\
\hline
3 & 0 [1], 3 [1], 6 [1]\\
\hline
4 & 0 [1], 4 [.980], 6 [.980], 8 [.997], 12 [1]\\
\hline
5 & 0 [1], 5 [.986], 8 [.983], 10 [.986], 12 [.979], 15 [.996], 20 [1]\\
\hline
6 & 0 [1], 6 [.989], 10 [.956], 12 [.770], 12[.745], 15 [.977], \\
 & 18 [.981], 18 [.240], 20 [.978], 24 [.996], 30 [1]\\
\hline
7 & 0 [1], 7 [.992], 12 [.931], 15 [.971], 18 [.952], 20 [.948], \\
 & 22 [.920], 24 [.970], 27 [.963], 30 [.979], 35 [.996], 42 [1]\\
\hline
8 & 0 [1], 8 [.993], 14 [.886], 18 [.959], 21 [.915], 24 [.960], \\
 & 26 [.917], 28 [.946], 28 [.072], 30 [.943], 32 [.917], 35 [.963], \\
 & 38 [.972], 42 [.980], 48 [.996], 56 [1] \\
\hline
9 & 0 [1], 9 [.994], 16 [.861], 21 [.926], 24 [.541], 24 [.854], \\
 & 28 [.944], 30 [.795], 30 [.388], 33 [.912], 35 [.937], 37 [.899], \\
 & 39 [.911], 42 [.074], 42 [.927], 44 [.917], 48 [.081], 48 [.958], \\
 & 51 [.978],  56 [.981], 63 [.996], 72 [1] \\
\hline
10 & 0 [1], 10 [.995], 18 [.848], 24 [.853], 28 [.934], 32 [.907], \\
 & 35 [.902], 38 [.872], 40 [.733], 40 [.607], 42 [.016], 42 [.577],  \\
 & 42 [.779], 45 [.894]\ldots 90 [1] 
\end{tabular}
\caption{For each number of bosons, $N$, the angular momenta, $L^*_N$,
at which the non-interacting composite fermion description predicts a
downward cusp are given, together with [in brackets] the overlaps of
the ansatz wavefunction (\protect\ref{eq:transform}) with the exact
groundstate wavefunction.  Where a given angular momentum appears more
than once for fixed $N$, the composite fermion model provides more
than one candidate groundstate.}
\label{table}
\end{table}



\begin{thebibliography}{10}
\bibitem{lanc}
These results were obtained from a combination of complete
diagonalisations and Lanczos techniques. The results are limited to
$L\leq 45$ for $N=10$.
\bibitem{fillingfraction}
The filling fraction, $\nu$, is defined as the ratio of the average
number density of particles, to the average number density of magnetic
flux quanta piercing the plane.
\end{thebibliography}
\end{document}